\documentclass[prl,aps,twocolumn,superscriptaddress,showpacs]{revtex4}
\usepackage{graphicx}

\begin{document}

\title{       Spin-Orbital Entanglement and Violation of
              the Goodenough-Kanamori Rules }

\author {     Andrzej M. Ole\'{s} }
\affiliation{ Max-Planck-Institut f\"ur Festk\"orperforschung,
              Heisenbergstrasse 1, D-70569 Stuttgart, Germany }
\affiliation{ Marian Smoluchowski Institute of Physics, Jagellonian
              University, Reymonta 4, PL-30059 Krak\'ow, Poland }

\author {     Peter Horsch }
\affiliation{ Max-Planck-Institut f\"ur Festk\"orperforschung,
              Heisenbergstrasse 1, D-70569 Stuttgart, Germany }

\author {     Louis Felix Feiner }
\affiliation{ Institute for Theoretical Physics, Utrecht University,
              Leuvenlaan 4, 3584 CC Utrecht}
\affiliation{ Philips Research Laboratories, High Tech Campus 4,
              5656 AE Eindhoven, The Netherlands }

\author {     Giniyat Khaliullin }
\affiliation{ Max-Planck-Institut f\"ur Festk\"orperforschung,
              Heisenbergstrasse 1, D-70569 Stuttgart, Germany }

\date{22 December 2005}

\begin{abstract}
We point out that large composite spin-orbital fluctuations in Mott
insulators with $t_{2g}$ orbital degeneracy are a manifestation of
quantum entanglement of spin and orbital variables. This results in
a dynamical nature of the spin superexchange interactions, which
fluctuate over positive and negative values, and leads to an
apparent violation of the Goodenough-Kanamori rules. \newline
[{\it Published in Phys. Rev. Lett. {\bf 96}, 147205 (2006).}]
\end{abstract}


\maketitle

%%%%%%%%%%%%%%%%%%%%%%%%%%%%%%%%%%%%%%%%%%%%%%%%%%%%%%%%%%%%%%%%%%%%%%%%
%%                      Goodenough-Kanamori rules
%%%%%%%%%%%%%%%%%%%%%%%%%%%%%%%%%%%%%%%%%%%%%%%%%%%%%%%%%%%%%%%%%%%%%%%%
Since the 1960's the magnetism of correlated Mott insulators like
transition-metal oxides has been understood by means of the
Goodenough-Kanamori (GK) rules \cite{Goode,Kan59}.
These state that if there is large overlap between partly occupied
orbitals at two magnetic ions, the superexchange interaction between
them is strongly antiferromagnetic (AF) because of the Pauli
principle, whereas overlap between partly occupied and unoccupied
orbitals gives weakly ferromagnetic (FM) interaction due to Hund's
exchange \cite{And63}. In the archetypical case of 180$^{\circ}$ bonds
through a single ligand ion this translates into a complementary
interdependence between spin order and orbital order \cite{Ima98}:
ferro orbital (FO) order supports strong AF spin order,
while alternating orbital (AO) order supports weak FM spin order.
The canonical example of this behavior is KCuF$_3$, where
weak FM (positive) spin correlations in the $ab$ planes and
strong AF (negative) correlations along the $c$ axis are accompanied by
AO order in the $ab$ planes and FO order along the $c$ axis.

%%%%%%%%%%%%%%%%%%%%%%%%%%%%%%%%%%%%%%%%%%%%%%%%%%%%%%%%%%%%%%%%%%%%%%%%
%%                    Typically the GK rules hold
%%%%%%%%%%%%%%%%%%%%%%%%%%%%%%%%%%%%%%%%%%%%%%%%%%%%%%%%%%%%%%%%%%%%%%%%
The GK rules (and extensions thereof \cite{footext}) have been extremely
successful in explaining the magnetic structure in a wide range of
materials. This may seem surprising because they presuppose that the
orbital occupation is static, whereas in recent years it has become
clear that if partly filled orbitals are degenerate, both spin and
orbital degrees of freedom should be considered as dynamic quantum
variables and be described by so-called spin-orbital models
\cite{Kug73,Tok00}. The GK rules work that well because in many
compounds a structural phase transition, driven by the Jahn-Teller (JT)
coupling of degenerate orbitals to the lattice, lifts the degeneracy
and fixes the orbital occupation well above the magnetic transition.
This happens typically for electrons in $e_g$ orbitals where large JT
distortions favor $C$-type orbital order, as in KCuF$_3$. However, for
$t_{2g}$ orbitals the JT coupling is rather weak, and recent experiments
in pseudocubic perovskite titanates \cite{Kei00} and vanadates
\cite{Ulr03} indeed indicate that the relevant orbitals 
{\it fluctuate\/},
and the conditions for applying the GK rules are not satisfied.

%%%%%%%%%%%%%%%%%%%%%%%%%%%%%%%%%%%%%%%%%%%%%%%%%%%%%%%%%%%%%%%%%%%%%%%%
%%                             this Letter
%%%%%%%%%%%%%%%%%%%%%%%%%%%%%%%%%%%%%%%%%%%%%%%%%%%%%%%%%%%%%%%%%%%%%%%%
In this Letter we investigate the magnetism of correlated insulators
in the case where classical static orbital order is absent. We will show 
that spins and orbitals then get entangled due to composite spin-orbital 
quantum fluctuations and that the familiar static GK rules are violated 
to the extent that even the signs of the magnetic interactions 
may fluctuate in time. To demonstrate this general feature in the most 
transparent way, we consider three different spin-orbital models for 
correlated insulators with 180$^{\circ}$ perovskite bonds between $d^1$, 
$d^2$ and $d^9$ ionic configurations, respectively, where the GK rules
definitely predict complementary signs of spin and orbital intersite 
correlations. The first two models are
derived for $t_{2g}$ electrons as in LaTiO$_3$ ($d^1$) and LaVO$_3$
($d^2$), where we demonstrate the violation of the GK rules, while the
third one is for $e_g$ holes as in KCuF$_3$ ($d^9$), in which the GK
rules are perfectly obeyed. This qualitative difference results from the
quantum nature of $t_{2g}$ orbitals which may form singlets, while $e_g$
orbitals behave more Ising-like and orbital singlets cannot form.

%%%%%%%%%%%%%%%%%%%%%%%%%%%%%%%%%%%%%%%%%%%%%%%%%%%%%%%%%%%%%%%%%%%%%%%%
%%                            som - general
%%%%%%%%%%%%%%%%%%%%%%%%%%%%%%%%%%%%%%%%%%%%%%%%%%%%%%%%%%%%%%%%%%%%%%%%
Superexchange may be regarded to arise from virtual excitations into
upper Hubbard bands, due to hopping with amplitude $t$, while low-energy
charge excitations are quenched by strong on-site Coulomb interaction
$U$. The resulting spin-orbital models take the generic form
\begin{equation}
\label{som}
{\cal H}=J\sum_{\gamma}\sum_{\langle ij\rangle\parallel\gamma}\left[
    \Big({\vec S}_i\cdot {\vec S}_j+S^2\Big){\hat J}_{ij}^{(\gamma)}
    + {\hat K}_{ij}^{(\gamma)}\right]+{\cal H}_{\rm orb},
\end{equation}
where $\gamma=a,b,c$ labels the cubic axes. The first term describes
the superexchange interactions ($J=4t^2/U$ is the superexchange
constant) between transition metal ions in the $d^n$ configuration with
spin $S$. The orbital operators ${\hat J}_{ij}^{(\gamma)}$ and
${\hat K}_{ij}^{(\gamma)}$ depend on Hund's exchange parameter
$\eta=J_H/U$, which determines the spectra of the virtual
$d^n_id^n_j\rightarrow d^{n+1}_id^{n-1}_j$ charge excitations.
In all three models considered here, for each axis $\gamma$ only two
orbital flavors are relevant, and ${\hat J}_{ij}^{(\gamma)}$ and
${\hat K}_{ij}^{(\gamma)}$ can be expressed in terms of pseudospin
$T=1/2$ operators $\vec{T}_i$ and $\vec{T}_j$.
Finally, ${\cal H}_{\rm orb}$ stands for the orbital interactions
(of strength $V$) induced by the coupling to the lattice
--- its form depends on the type of orbitals ($t_{2g}$ or $e_g$).

%%%%%%%%%%%%%%%%%%%%%%%%%%%%%%%%%%%%%%%%%%%%%%%%%%%%%%%%%%%%%%%%%%%%%%%%
%%                 coupling of spins and t2g orbitals
%%%%%%%%%%%%%%%%%%%%%%%%%%%%%%%%%%%%%%%%%%%%%%%%%%%%%%%%%%%%%%%%%%%%%%%%
For the $t_{2g}$ systems we will consider chains along the $c$ axis,
where only two ($yz$ and $zx$) orbital flavors are active, i.e.
participate in the hopping. We assume the idealized case where these
two orbitals contain one electron per site, which implies that the
third ($xy$) orbital is empty in the $d^1$ model and filled by one
electron in the $d^2$ model. The orbital operators,
${\hat J}_{ij}^{(c)}(d^1)$ \cite{Kha00,Kha03}
and ${\hat J}_{ij}^{(c)}(d^2)$ \cite{Kha01},
describing the coupling between the $S=1/2$ spins of the Ti$^{3+}$
($d^1$) ions in cubic titanates and that between the $S=1$ spins of the
V$^{3+}$ ($d^2$) ions in cubic vanadates, respectively, reduce in the
absence of Hund's coupling to an SU(2)-symmetric expression
$\propto ({\vec T}_i\cdot {\vec T}_j+\textstyle{\frac{1}{4}})$,
which may take both positive and negative values. Note also that the
superexchange [see Eq. (\ref{som})] thus contains interactions like
$(S_i^+ T_i^-)(S_j^- T_j^+) + (S_i^- T_i^+)(S_j^+ T_j^-)$,
which generate {\it simultaneous\/} fluctuations of spins and orbitals
described by the composite operators $Q_i^+ \equiv S_i^+ T_i^-$ etc.
At finite $\eta$ both ${\hat J}_{ij}^{(c)}(d^1)$ and
${\hat J}_{ij}^{(c)}(d^2)$ also contain
\begin{equation}
{\vec T}_i\otimes{\vec T}_j=\textstyle{\frac{1}{2}}
\big(T_i^+ T_j^+ + T_i^- T_j^- \big) + T_i^z T_j^z.
\label{tau++}
\end{equation}
This operator appears because double occupancy of either active
($yz$ or $zx$) orbital is not an eigenstate of the on-site Coulomb
interaction. Consequently, the total $T$ and $T^z$ quantum numbers
are not conserved and orbital fluctuations are amplified. Finally,
GdFeO$_3$-type distortions induce orbital interactions $\propto
-VT_i^zT_j^z$ favoring FO order along the $c$ axis \cite{Miz99}.

%%%%%%%%%%%%%%%%%%%%%%%%%%%%%%%%%%%%%%%%%%%%%%%%%%%%%%%%%%%%%%%%%%%%%%%%
%%                               d9 model
%%%%%%%%%%%%%%%%%%%%%%%%%%%%%%%%%%%%%%%%%%%%%%%%%%%%%%%%%%%%%%%%%%%%%%%%
In the $e_g$ system there are two orbital flavors ($3z^2-r^2$ and
$x^2-y^2$), and for each axis a different linear combination of them
is active ($3x^2-r^2$ along $a$, $3y^2-r^2$ along $b$, and $3z^2-r^2$
along $c$). Thus the superexchange ${\hat J}_{ij}^{(\gamma)}(d^9)$
between the $S=1/2$ spins at the Cu$^{2+}$ $(d^9)$ ions in KCuF$_3$ is
expressed \cite{Fei97} in terms of axis-dependent orbital operators
$T^{(a,b)}_i=-\textstyle{\frac{1}{4}}
                 (\sigma_i^z\mp\sqrt{3}\sigma_i^x)$ and
$T^{(c)}_i  = \textstyle{\frac{1}{2}}\sigma_i^z$,
given by Pauli matrices $\sigma_i^x$ and $\sigma_i^z$.
In the absence of Hund's coupling
${\hat J}_{ij}^{(\gamma)}(d^9)=
(T_i^{(\gamma)}-\textstyle{\frac{1}{2}})
(T_j^{(\gamma)}-\textstyle{\frac{1}{2}})$,
which, in sharp contrast to the $t_{2g}$ case above, is never negative
owing to only a single orbital being active along each axis.
In formal terms, ${\hat J}_{ij}^{(\gamma)}(d^9)$ is not SU(2)-symmetric,
and thus orbital singlets are not formed.
The Ising-like form of ${\hat J}_{ij}^{(\gamma)}(d^9)$ makes
the $d^9$ model look more classical than the $t_{2g}$ models,
but spin-orbital dynamics is still promoted as the orbital flavor is
not conserved \cite{Fei05}. Finally, the JT ligand distortions around
Cu$^{2+}$ ions lead to orbital interactions
$\propto V T_i^{(\gamma)} T_j^{(\gamma)}$ that favor AO order.

%%%%%%%%%%%%%%%%%%%%%%%%%%%%%%%%%%%%%%%%%%%%%%%%%%%%%%%%%%%%%%%%%%%%%%%%
%%                         correlation functions
%%%%%%%%%%%%%%%%%%%%%%%%%%%%%%%%%%%%%%%%%%%%%%%%%%%%%%%%%%%%%%%%%%%%%%%%
We investigated intersite spin, orbital and composite spin-orbital
correlations in the above spin-orbital models. To make the results
comparable in all cases, we use
\begin{equation}
\label{ss}
S_{ij}=\langle{\vec S}_i\cdot {\vec S}_j\rangle/(2S)^2
\end{equation}
for the spin correlations. The orbital and spin-orbital correlations are
defined for the $t_{2g}$ ($d^1$ and $d^2$) models as
\begin{eqnarray}
\label{tt}
T_{ij}^{(t)}\!\!&=&\!\big\langle{\vec T}_i\cdot{\vec T}_j\big\rangle,\\
\label{ct}
C_{ij}^{(t)}\!\!&=&\!\!
\big[\big\langle({\vec S}_i\!\cdot\!{\vec S}_j)
     ({\vec T}_i\!\cdot\!{\vec T}_j)\big\rangle\!
     -\!\big\langle{\vec S}_i\!\cdot\!{\vec S}_j\big\rangle
      \big\langle{\vec T}_i\!\cdot\!{\vec T}_j\big\rangle\big]/(2S)^2,
\end{eqnarray}
while for the $e_g$ ($d^9$) model
\begin{eqnarray}
\label{te}
T_{ij}^{(e)}\!\!&=&\!
\big\langle{T}_i{T}_j-\textstyle{\frac{1}{2}}({T}_i+{T}_j)
\big\rangle^{(\gamma)},        \\
\label{ce}
C_{ij}^{(e)}\!\!&=&\!
\big\langle({\vec S}_i\!\cdot\!{\vec S}_j)[{T}_i{T}_j
    -\textstyle{\frac{1}{2}}({T}_i+{T}_j)]\big\rangle^{(\gamma)}\!
    -S_{ij}^{}T_{ij}^{(e)}.
\end{eqnarray}
These definitions of $C_{ij}^{(t,e)}$ are dictated by the structure
of the spin-orbital superexchange in the $J_H\to 0$ limit.

\begin{figure}[b!]
\includegraphics[width=6.8cm]{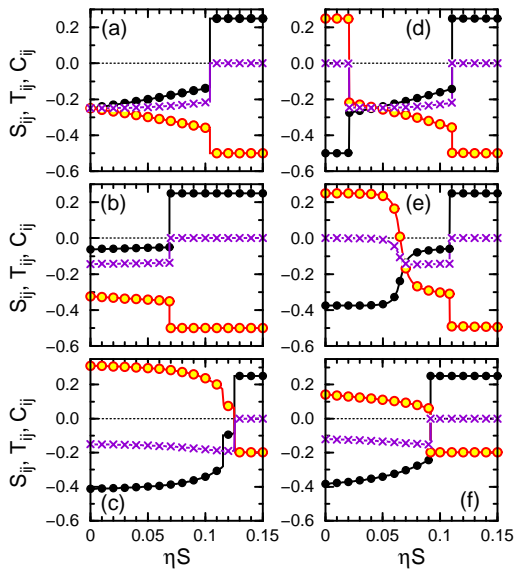}
\vskip -.2cm
\caption{(Color online)
Intersite spin $S_{ij}$ (filled circles),
orbital $T_{ij}^{(t,e)}$ (empty circles),
and composite spin-orbital $C_{ij}^{(t,e)}$ (crosses) correlations
as functions of Hund's exchange $\eta S$, for $V=0$ (left) and for
$V=J$ (right) for:
(a,d) $d^1$ model,
(b,e) $d^2$ model, and
(c,f) $d^9$ model.
}
\label{fig:t0}
\end{figure}

%%%%%%%%%%%%%%%%%%%%%%%%%%%%%%%%%%%%%%%%%%%%%%%%%%%%%%%%%%%%%%%%%%%%%%%%
%%                                 d1
%%%%%%%%%%%%%%%%%%%%%%%%%%%%%%%%%%%%%%%%%%%%%%%%%%%%%%%%%%%%%%%%%%%%%%%%
We have solved both $t_{2g}$ models, $d^1$ and $d^2$, on four-site
chains along the $c$ axis using periodic boundary conditions, and we 
find that nontrivial spin-orbital dynamics strongly influences the 
intersite correlations. First we consider $V=0$, i.e. the purely 
electronic (superexchange) spin-orbital models. In the titanate $d^1$ 
case one recovers the SU(4) model \cite{Li98} in the limit of $\eta=0$, 
with robust SU(4) singlet correlations \cite{Bos01}. Indeed, in the 
four-site chain all intersite correlations are 
{\it identical and negative\/},
$S_{ij}=T_{ij}^{(t)}=C_{ij}^{(t)}=-0.25$ [Fig. \ref{fig:t0}(a)].
As expected, this value is somewhat lower than $-0.215$ obtained for
the infinite SU(4) chain \cite{Fri99}. At finite $\eta$ one finds
$T_{ij}^{(t)}<C_{ij}^{(t)}<S_{ij}<0$ as long as the spin singlet ($S=0$)
ground state persists, i.e. for $\eta\lesssim 0.21$, and the GK rules,
which imply that the signs of $S_{ij}$ and $T_{ij}^{(t)}$ are different
(spin and orbital correlations are complementary) are violated.
Apparently the composite spin-orbital correlations $C_{ij}^{(t)}<0$
dominate and cannot be determined from $S_{ij}$ and $T_{ij}^{(t)}$ by
mean-field (MF) decoupling, so the spin and orbital variables are
{\it entangled\/}, similar to entanglement in pure spin models
\cite{Fan04}. In fact, the values of the correlations indicate that the 
wavefunction on a bond $\langle ij\rangle$ is close to a singlet of the 
(total) composite quasi-spin ${\vec Q}_i + {\vec Q}_j$, equivalent to a 
linear combination of (spin-singlet/orbital-triplet) and
(spin-triplet/orbital-singlet).

%%%%%%%%%%%%%%%%%%%%%%%%%%%%%%%%%%%%%%%%%%%%%%%%%%%%%%%%%%%%%%%%%%%%%%%%
%%                     d2 + Conventional picture
%%%%%%%%%%%%%%%%%%%%%%%%%%%%%%%%%%%%%%%%%%%%%%%%%%%%%%%%%%%%%%%%%%%%%%%%
The vanadate $d^2$ model behaves similarly, with all three $S_{ij}$,
$T_{ij}^{(t)}$ and $C_{ij}^{(t)}$ correlations being negative in the
spin-singlet ($S=0$) orbital-disordered phase, stable for
$\eta\lesssim 0.07$ [Fig. \ref{fig:t0}(b)]. Here the spin correlations
are weakly AF ($S_{ij}\simeq -0.05$), and AF and FM bonds compete,
promoting a dimerized state \cite{Zha02}.
--- For both ($d^1$ and $d^2$) models the conventional picture is
restored at large Hund's coupling, which stabilizes the FM ground states
(at $\eta\gtrsim 0.21$ for $d^1$, and at $\eta\gtrsim 0.07$ for $d^2$).
Here the spin-orbital correlations decouple ($C_{ij}^{(t)}=0$) and
the GK rules are perfectly obeyed, with positive $S_{ij}=0.25$ (FM)
and negative $T_{ij}^{(t)}=-0.5$ (AO) correlations.

%%%%%%%%%%%%%%%%%%%%%%%%%%%%%%%%%%%%%%%%%%%%%%%%%%%%%%%%%%%%%%%%%%%%%%%%
%%                              figure 1
%%%%%%%%%%%%%%%%%%%%%%%%%%%%%%%%%%%%%%%%%%%%%%%%%%%%%%%%%%%%%%%%%%%%%%%%

%%%%%%%%%%%%%%%%%%%%%%%%%%%%%%%%%%%%%%%%%%%%%%%%%%%%%%%%%%%%%%%%%%%%%%%%
%%                         correlations in d9
%%%%%%%%%%%%%%%%%%%%%%%%%%%%%%%%%%%%%%%%%%%%%%%%%%%%%%%%%%%%%%%%%%%%%%%%
The $d^9$ model shows completely different behavior. Considering a
four-site plaquette in the $ab$ plane, one finds that the conventional
spin-orbital interrelation (AF/FO or FM/AO) is a robust property of the
model at any value of Hund's coupling. For small $\eta\lesssim 0.25$,
FO correlations $T_{ij}^{(e)}>0$ are accompanied by strong AF spin 
correlations, $S_{ij}<0$, and this changes into the opposite at large 
$\eta$ [Fig. \ref{fig:t0}(c)], just as one would expect from the
GK rules. Reflecting this situation, the composite spin-orbital
correlations $C_{ij}^{(e)}$ are weaker than the spin correlations
$S_{ij}$ and the orbital correlations $T_{ij}^{(e)}$.
This permits spin-orbital separation in the ground state,
and corrections to this picture are only perturbative \cite{Kha97}.

%%%%%%%%%%%%%%%%%%%%%%%%%%%%%%%%%%%%%%%%%%%%%%%%%%%%%%%%%%%%%%%%%%%%%%%%
%%                              figure 2
%%%%%%%%%%%%%%%%%%%%%%%%%%%%%%%%%%%%%%%%%%%%%%%%%%%%%%%%%%%%%%%%%%%%%%%%
\begin{figure}[b!]
\includegraphics[width=6.8cm]{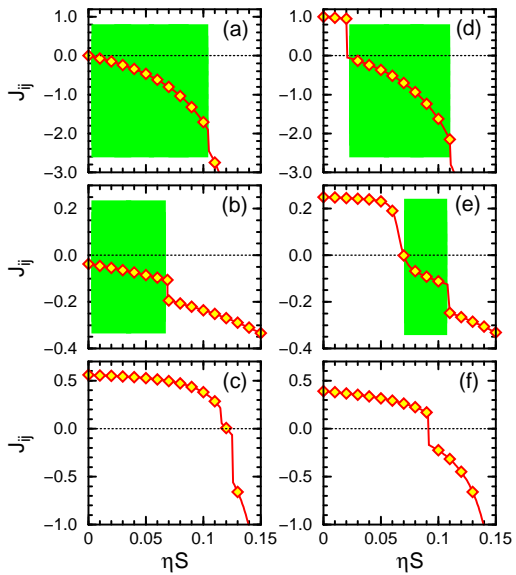}
\vskip -.2cm
\caption{(Color online)
Spin exchange constants $J_{ij}$ at $V=0$ (left) and at $V=J$ (right)
as functions of Hund's exchange $\eta S$ for
(a), (d) $d^1$ model;
(b), (e) $d^2$ model;
(c), (f) $d^9$ model.
In the shaded regions in (a), (b), (d), (e), $J_{ij}$ is negative 
(FM) and yet the spin correlations are AF, $S_{ij}<0$ 
(see Fig. \ref{fig:t0}).
}
\label{fig:allj}
\end{figure}

%%%%%%%%%%%%%%%%%%%%%%%%%%%%%%%%%%%%%%%%%%%%%%%%%%%%%%%%%%%%%%%%%%%%%%%%
%%                           lattice effects
%%%%%%%%%%%%%%%%%%%%%%%%%%%%%%%%%%%%%%%%%%%%%%%%%%%%%%%%%%%%%%%%%%%%%%%%
Next we consider finite $V$, where one expects that the coupling to the
lattice could suppress the orbital fluctuations and cure the apparent
violation of the GK rules in the $t_{2g}$ models. Indeed, at small
$\eta$ finite $V$ induces orbital order and so stabilizes the AF/FO
phase [Figs. \ref{fig:t0}(d) and \ref{fig:t0}(e)], composite
spin-orbital fluctuations are suppressed and the GK rules are restored.
Already infinitesimal interaction $V>0$ removes the SU(4) symmetry of
the $d^1$ model at $\eta=0$ by an Ising-like orbital anisotropy.
However,
for sufficiently large Hund's exchange $\eta$ the spin-singlet phase
survives (unless $V \gg J$, i.e. orbital interactions much stronger
than the superexchange). At $V=J$ one thus finds three magnetic phases
in the $d^1$ ($d^2$) model [Figs. \ref{fig:t0}(d) and \ref{fig:t0}(e)]:
  (i) AF/FO order \cite{noteaf} in the range of
      $\eta \lesssim 0.04$ ($\eta\lesssim 0.06$);
 (ii) an intermediate orbital-disordered phase with negative spin,
      orbital and composite spin-orbital correlations
      of about equal strength, and
(iii) FM/AO order for $\eta \gtrsim 0.22$ ($\eta \gtrsim 0.11$).
The first two are separated by an orbital transition within the
spin-singlet phase. Notably, the GK rules are perfectly obeyed in
phases (i) and (iii) \cite{notev9}, while again they do not apply in
the intermediate phase (ii), which is moved now to a more realistic
regime of larger $\eta S$.

In the $d^9$ case finite $V$ only stabilizes the large-$\eta$ phase
with FM/AO order at the expense of the small-$\eta$ AF/FO phase, but
the behavior of the model is not changed qualitatively [compare Figs.
\ref{fig:t0}(c) and \ref{fig:t0}(f)].
We emphasize that the different behavior of $t_{2g}$ and $e_g$ systems
is intrinsic, i.e. has its origin in different spin-orbital physics
generated by the electronic superexchange interactions. In particular,
it is not caused by being affected differently by coupling to the
lattice (i.e., by finite $V$).

%%%%%%%%%%%%%%%%%%%%%%%%%%%%%%%%%%%%%%%%%%%%%%%%%%%%%%%%%%%%%%%%%%%%%%%%
%%                          exchange constants
%%                          fluctuations of <J>
%%%%%%%%%%%%%%%%%%%%%%%%%%%%%%%%%%%%%%%%%%%%%%%%%%%%%%%%%%%%%%%%%%%%%%%%
Further evidence that the GK rules do not directly apply in $t_{2g}$
systems follows from the spin exchange constants
$J_{ij}\equiv\langle \hat{J}_{ij}^{(\gamma)} \rangle$,
the expectation value being taken over the orbital variables.
One finds that in the orbital-disordered phase formally FM interaction
$J_{ij}<0$ is in fact, both for $V=0$ and for finite $V$, accompanied
by AF spin correlations [Figs. \ref{fig:allj}(a,d) and 
\ref{fig:allj}(b,e)], whereas in the $e_g$ case the spin correlations 
follow the sign of $J_{ij}$ for all values of $\eta$ 
[Fig. \ref{fig:allj}(c,f)]. This remarkable difference between $t_{2g}$ 
and $e_g$ systems is due to composite spin-orbital fluctuations, which 
are responsible for {\it `dynamical'\/} exchange constants 
$\hat{J}_{ij}^{(\gamma)}$ in the former case, which exhibit large 
fluctuations, measured by
$\delta J=(\langle (\hat{J}_{ij}^{(\gamma)})^2\rangle-J_{ij}^2)^{1/2}$
\cite{Kha03}, as we illustrate here at $\eta=0$. While the average spin 
exchange constant is small in both $t_{2g}$ models ($J_{ij} \simeq 0$ 
for $d^1$, $J_{ij} \simeq -0.04$ for $d^2$), $\hat{J}_{ij}^{(\gamma)}$ 
fluctuates widely over both positive and negative values. In the $d^1$ 
case the fluctuations between ($S=0$/$T=1$) and ($S=1$/$T=0$) bond 
states are so large that $\delta J=1$ !
They survive even quite far from the high-symmetry SU(4)
point (at $\eta>0.1$). Also in the $d^2$ model the orbital bond
correlations change dynamically from singlet to triplet \cite{Kha01},
resulting in $\delta J>|J_{ij}|$, with
$\delta J=\frac{1}{4}\{1-(2T_{ij}+\frac{1}{2})^2\}^{1/2}\simeq 0.247$.
In contrast, the more classical behavior in the $d^9$ case is confirmed
by
$\delta J<J_{ij}$, as $J_{ij} \simeq 0.56$
and
$\delta J=\frac{1}{2}\{1-(2T_{ij}-\frac{1}{2})^2\}^{1/2}\simeq 0.50$.

%%%%%%%%%%%%%%%%%%%%%%%%%%%%%%%%%%%%%%%%%%%%%%%%%%%%%%%%%%%%%%%%%%%%%%%%
%%                            Table: energies
%%%%%%%%%%%%%%%%%%%%%%%%%%%%%%%%%%%%%%%%%%%%%%%%%%%%%%%%%%%%%%%%%%%%%%%%
\begin{table}[t!]
\caption{
Energies per site: exact $E_0$ and MF $E_{\rm MF}$ for the spin-singlet
phases in the three spin-orbital models, obtained with four-site
clusters. All energies and $V$ in units of $J$.
}
\begin{ruledtabular}
\begin{tabular}{cccccccc}
         &       & \multicolumn{2}{c}{$d^1$ model} &
                   \multicolumn{2}{c}{$d^2$ model} &
                   \multicolumn{2}{c}{$d^9$ model} \cr
$\eta S$ & $V$ & $E_0$ & $E_{\rm MF}$ & $E_0$ & $E_{\rm MF}$
                                      & $E_0$ & $E_{\rm MF}$ \cr
\colrule
 0.0  & 0.0 & -0.500 &  0.0   & -0.316 & -0.028 & -0.594 & -0.443 \cr
 0.06 & 0.0 & -0.655 & -0.006 & -0.388 & -0.112 & -0.634 & -0.472 \cr
 0.07 & 1.0 & -0.607 &  0.097 & -0.311 & -0.027 & -0.633 & -0.487 \cr
\end{tabular}
\end{ruledtabular}
\label{tab:e0}
\end{table}
%%%%%%%%%%%%%%%%%%%%%%%%%%%%%%%%%%%%%%%%%%%%%%%%%%%%%%%%%%%%%%%%%%%%%%%%

%%%%%%%%%%%%%%%%%%%%%%%%%%%%%%%%%%%%%%%%%%%%%%%%%%%%%%%%%%%%%%%%%%%%%%%%
%%                        discussion of energies
%%%%%%%%%%%%%%%%%%%%%%%%%%%%%%%%%%%%%%%%%%%%%%%%%%%%%%%%%%%%%%%%%%%%%%%%
When quantum entanglement occurs, the ground state energy $E_0$ cannot
be estimated reliably by MF decoupling of composite correlations
(i.e., with the assumption $C_{ij}^{(t,e)}=0$). The corrections beyond 
the MF energy $E_{\rm MF}$ are largest in the $d^1$ case and remain 
significant in the $d^2$ model (Table \ref{tab:e0}), but are much less 
pronounced in the $d^9$ model, even at $V=0$. Only when such corrections 
disappear, orbitals disentangle from spins and can be analyzed 
separately \cite{vdB04}, or spin states can be treated for fixed orbital 
order according to the (static) GK rules.

%%%%%%%%%%%%%%%%%%%%%%%%%%%%%%%%%%%%%%%%%%%%%%%%%%%%%%%%%%%%%%%%%%%%%%%%
%%                      phase transition in d^2 model
%%%%%%%%%%%%%%%%%%%%%%%%%%%%%%%%%%%%%%%%%%%%%%%%%%%%%%%%%%%%%%%%%%%%%%%%
We further notice that the $d^2$ model exhibits an interesting property
related to the nature of transitions between different phases.
Namely, the ground state at $V=J$ is a nondegenerate spin-singlet for
$0<\eta \lesssim 0.11$, while the orbital quantum numbers change
gradually from
$\langle T \rangle \simeq 2$ and $\langle T^z \rangle \simeq\pm 2$ to
$\langle T \rangle \simeq 0$ and $\langle T^z \rangle \simeq 0$
in the crossover regime of $\eta\simeq 0.06$ [see Fig. \ref{fig:t0}(e)].
We have verified that when the orbital terms $\propto T_i^+ T_j^+$
are neglected, i.e., if Eq. (\ref{tau++}) is replaced by an Ising-like
term $T_i^z T_j^z$,
a sharp transition occurs instead (from the doubly degenerate FO state
with $T^z=\pm 2$ to a disordered state with $T=1$, $T^z=0$),
consistent with abrupt transitions found before for an infinite chain
\cite{Kaw04}. Therefore, we anticipate that the $T_i^+ T_j^+$ terms
induce a continuous orbital phase transition in the thermodynamic limit.

%%%%%%%%%%%%%%%%%%%%%%%%%%%%%%%%%%%%%%%%%%%%%%%%%%%%%%%%%%%%%%%%%%%%%%%%
%%                          realistic compounds
%%%%%%%%%%%%%%%%%%%%%%%%%%%%%%%%%%%%%%%%%%%%%%%%%%%%%%%%%%%%%%%%%%%%%%%%
We emphasize that composite spin-orbital fluctuations and dynamical
exchange constants will control, for realistic parameters,
the behavior of titanates and vanadates. In fact, the idea that
SU(4)-like fluctuations dominate in the ground state has been put
forward to understand the unusual properties of LaTiO$_3$ \cite{Kha00}
and the possible quantum critical point in the titanate phase diagram
\cite{Kha03,Kha05}. Such fluctuations also drive $C$-AF spin order in
LaVO$_3$ \cite{Kha01} and spin-orbital dimerization in YVO$_3$
\cite{Ulr03,Kaw04}.

%%%%%%%%%%%%%%%%%%%%%%%%%%%%%%%%%%%%%%%%%%%%%%%%%%%%%%%%%%%%%%%%%%%%%%%%
%%                              summary
%%%%%%%%%%%%%%%%%%%%%%%%%%%%%%%%%%%%%%%%%%%%%%%%%%%%%%%%%%%%%%%%%%%%%%%%
Summarizing, in correlated insulators with partly filled $t_{2g}$
shells, orbitals and spins are entangled, and average spin and orbital
correlations are typically in conflict with the (static) GK rules.
These rules should then instead be understood in terms of dynamical
spin and orbital correlations that are complementary to each other,
and both configurations --- (orbital-singlet/spin-triplet) and
(orbital-triplet/spin-singlet) --- are entangled in the ground state.
It remains both an experimental and theoretical challenge to investigate
the physical consequences of spin-orbital entanglement in real systems.

%%%%%%%%%%%%%%%%%%%%%%%%%%%%%%%%%%%%%%%%%%%%%%%%%%%%%%%%%%%%%%%%%%%%%%%%
%%                          Acknowledgments
%%%%%%%%%%%%%%%%%%%%%%%%%%%%%%%%%%%%%%%%%%%%%%%%%%%%%%%%%%%%%%%%%%%%%%%%
This work was supported by the Polish Ministry of Science and Education
Project No.~1 P03B 068 26.

\vskip -.1cm
%%%%%%%%%%%%%%%%%%%%%%%%%%%%%%%%%%%%%%%%%%%%%%%%%%%%%%%%%%%%%%%%%%%%%%%%
%%
%%                             REFERENCES
%%
%%%%%%%%%%%%%%%%%%%%%%%%%%%%%%%%%%%%%%%%%%%%%%%%%%%%%%%%%%%%%%%%%%%%%%%%


\begin{thebibliography}{}
\bibitem{noteaf} AF order occurs when the cluster is embedded by
                 MF terms and solved self-consistently in higher 
		 dimension.
\end{thebibliography}
\end{document}